\documentclass[12pt,preprint]{aastex}
\usepackage{epsf}
\usepackage{lscape}

\usepackage{graphicx}
\usepackage{natbib}
\begin{document}
\title{A Redshift Survey of the Strong Lensing Cluster Abell 383}

\author {Margaret J. Geller} 
\affil{Smithsonian Astrophysical Observatory,
\\ 60 Garden St., Cambridge, MA 02138}
\email{mgeller@cfa.harvard.edu}
\author {Ho Seong Hwang} 
\affil{Smithsonian Astrophysical Observatory,
\\ 60 Garden St., Cambridge, MA 02138}
\email{hhwang@cfa.harvard.edu}
\author{Antonaldo Diaferio}
\affil{Dipartimento di Fisica,
\\Universit\`a  di Torino, via P. Giuria 1, 10125 Torino, Italy}
\affil{ INFN, Sezione di Torino, via P. Giuria 1, 
\\10125 Torino, Italy}
\email{diaferio@ph.unito.it}
\author {Michael J. Kurtz} 
\affil{Smithsonian Astrophysical Observatory,
\\ 60 Garden St., Cambridge, MA 02138}
\email{mkurtz@cfa.harvard.edu}
\author{Dan Coe}
\affil{Space Telescope Science Institute, Baltimore, MD}
\email {DCoe@STScI.edu}
\author{Kenneth J. Rines}
\affil {Department of Physics \& Astronomy, Western Washington University, Bellingham, WA 98225}
\email{kenneth.rines@wwu.edu}

\begin{abstract}
Abell 383 is a famous rich cluster (z = 0.1887) imaged extensively as a basis for intensive strong and weak lensing studies.
Nonetheless there are few spectroscopic observations. We enable dynamical analyses by measuring  
2360 new redshifts for galaxies with r$_{petro} \leq 20.5$ and within 50$^\prime$ of the BCG (Brightest Cluster Galaxy: R.A.$_{2000} = 42.014125^\circ$, Decl$_{2000} = -03.529228^\circ$). We apply the caustic technique to identify
275 cluster members within 7$h^{-1}$ Mpc of the hierarchical cluster center. The BCG lies within
$-11 \pm 110$ km s$^{-1}$ and 21 $\pm 56 h^{-1}$ kpc of the hierarchical cluster center; the velocity dispersion profile of the BCG appears to be an extension of the velocity dispersion profile based on cluster members. The distribution of cluster members on the sky corresponds impressively with the weak lensing contours of Okabe et al. (2010) especially when the impact of foreground and background structure is included. The values of R$_{200}$ = $1.22\pm 0.01 h^{-1}$ Mpc and M$_{200}$ = $(5.07 \pm 0.09 )\times 10^{14} h^{-1}$ M$_\odot$
obtained by application of the caustic technique agree well with recent completely independent lensing measures. The caustic estimate extends direct measurement of the cluster mass profile to a radius of $\sim 5 h^{-1}$ Mpc.
\end{abstract}

\section {Introduction}

Massive clusters of galaxies are cornerstones for the determination of mass profiles in systems of galaxies. The most luminous, apparently relaxed systems serve as particularly important probes of
the matter distribution on scales from $\sim 40h^{-1}$ kpc to 5$h^{-1}$ Mpc.

The apparently relaxed, x-ray luminous cluster, A383 (z =0.1887) is a prototypical system (Ebeling et al. 1996; Allen et al. 2008) included in the CLASH program (Postman et al. 2012); several studies
address  strong and weak lensing (and a combination of these two methods)
measurements of the mass profile of A383 (Smith et al. 2001; Bardeau et al. 2007; Hoekstra 2007; Sand et al. 2008; Okabe et al. 2010; Newman et al. 2011;
Huang et al. 2011; Zitrin et al. 2011). Richard et al. (2011) and Zitrin et al. (2012) discuss the effectiveness of A383 in lensing distant background galaxies and clusters of galaxies, respectively.
Remarkably, in spite of the attention to this system, there is no substantial spectroscopic survey of the cluster and its environment.
We remedy this situation with an intensive  MMT Hectospec (Fabricant et al. 2005) redshift survey that includes
new redshifts for  2360 galaxies within
a $\sim 2^\circ$ field centered on A383. We identify 275 cluster members
within 7$h^{-1}$ Mpc of the cluster center.

The strong and weak lensing analyses of A383 yield a host of measures of the cluster mass distribution on scales ranging
from $\sim 30 h^{-1}$ kpc to $\sim 1h^{-1}$ Mpc. The many gravitationally lensed features in the cluster core include a giant arc, radial arcs within the halo of the brightest cluster galaxy, and many smaller arcs and arclets. This abundance of features enables detailed modeling of the matter distribution in the cluster center (Smith et al. (2001); 
Sand et al. (2008); Newman et al. (2011); Zitrin et al. (2011)). Combining strong lensing constraints and x-ray observations enables more complex modeling that constrains triaxility, principal axes orientation and non-thermal pressure (Morandi \& Limousin 2012). 


For A383, there are many independent weak lensing measurements of M$_{200}$, the cluster mass within a 
radius R$_{200}$ enclosing an average density 200 times the critical density. The estimates of 
Bardeau et al. (2007), Hoekstra et al. (2007), and Okabe et al. (2010) span the range (2.3 - 3.1)$\times 10^{14} h^{-1}$ M$_\odot$. More recent studies by Huang et al. (2011), Zitrin et al. (2011), and Newman et al. (2013a) favor a somewhat larger mass of $\sim 4.5 \times$10$^{14} h^{-1}$M$_\odot$.
An analysis of Chandra data by
Schmidt \& Allen (2007) favors the larger mass. Here we focus on these larger scale mass estimates and derive a completely independent M$_{200}$ from the redshift survey and compare it with the recent weak lensing results.

We describe the extensive Hectospec redshift survey of the cluster and its surroundings in Section \ref {data}. In Section \ref {members}
we use the caustic technique to determine the cluster membership.
We use the members  together with the velocity dispersion profile for the brightest cluster galaxy (BCG) from Newman et al. (2011; 2013a) to construct a velocity dispersion profile for the cluster on scales from $\sim 20 h^{-1}$ kpc to 5$h^{-1}$ Mpc (Section \ref{vprofile}). In Section \ref{weakmap} we compare the member distribution on the sky with weak lensing map of Okabe et al. (2010). We also discuss foreground and background structures and their possible impact on the weak lensing results.  In Section \ref{massprof} we compare the mass profile derived from the redshift survey by application of the caustic technique with a set of weak lensing mass profiles. 

We adopt H$_0$ = 100 $h$ km s$^{-1}$ Mpc$^{-1}$, $\Omega_\Lambda$ = 0.7 and $\Omega_m$ = 0.3 throughout.
All quoted errors in measured quantities are 1$\sigma$.
\section {Data}
\label {data}

Strong- and weak-lensing studies of A383 are based on abundant, high quality imaging data from CFHT, Subaru, and HST. Deep ground-based uBVRIz images taken with MEGACAM and SUPRIMECAM on the CFHT and Subaru, respectively, are the basis for the weak-lensing study by Huang et al. (2011). Zitrin et al. (2011) base their strong-lensing analysis on 16-band HST imaging with the WFC3 UVIS and IR cameras and the ACS WFC. A383 also lies within the footprint of DR10 of the Sloan Digital Sky Survey (SDSS: Ahn  et al. 2013). We summarize our use of the SDSS photometric data in Section \ref {photometry}.

In contrast with the abundant imaging data, there is very little published spectroscopy. Newman et al
(2011) measured a velocity dispersion profile for the brightest cluster galaxy. There is also some spectroscopy for gravitationally-lensed arcs (Smith et al. 2001; Sand et al. 2008; Newman et al. 2011; Zitrin et al. 2011).  In Section \ref {spectroscopy}
we describe our Hectospec redshift survey of A383 and its surroundings. 

\subsection {Photometric Data}
\label {photometry}

We used SDSS photometry from DR8 (Aihara et al. 2011) to select objects for spectroscopic observations. We chose 
objects with $r_{fiber} \leq 21$ and within 50$^\prime$ of the brightest cluster galaxy (BCG). This selection of high surface brightness objects enables relatively short integrations even in gray time or with substantial cirrus.
We weighted the spectroscopic targets according to their angular separation from the BCG 
(R.A.$_{2000} = 42.014125^\circ$, Decl$_{2000} = -03.529228^\circ$). We observed an essentially uniformly complete sample of galaxies with r$_{petro} \leq 20.5$ and within 25$^{\prime}$ of the BCG.

\subsection {Spectroscopy}
\label {spectroscopy}

The Hectospec instrument (Fabricant et al. 2005)
mounted on the MMT 6.5m telescope is an ideal instrument
for studying clusters and their infall regions at moderate redshift.
Hectospec is a multiobject fiber-fed spectrograph
with 300 fibers deployable over a circular field-of-view
with a diameter of 1$^\circ$. The spectra cover the wavelength range
3500 - 9150 $\AA$ at a resolution of 6.2 $\AA$ FWHM.

The galaxy targets are relatively bright for Hectospec
spectroscopy. We obtained high-quality spectra with
3x20-minute exposures even under suboptimal observing
conditions (e.g., poor seeing, thin clouds). 

After processing and reducing the spectra, we used
the IRAF package rvsao (Kurtz \& Mink 1998) to crosscorrelate
the spectra with galaxy templates assembled
from previous Hectospec observations. During the
pipeline processing, spectral fits are assigned a quality
flag of ``Q'' for high-quality redshifts, ``?'' for marginal cases, and ``X'' for poor fits.
We use only the spectra assigned ``Q''.

\begin{figure}[htb]
\centerline{\includegraphics[width=7.0in]{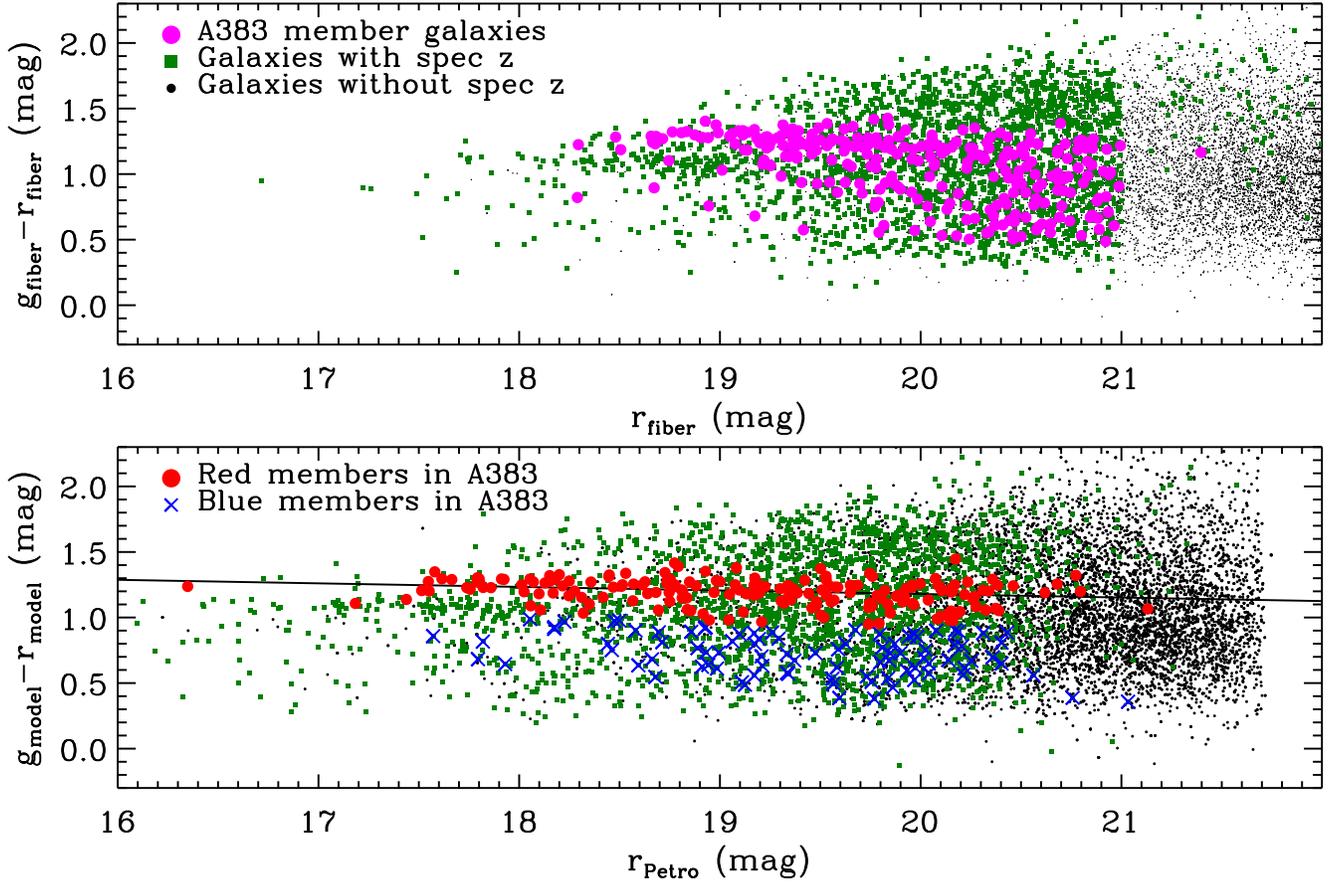}}
\vskip -2ex
\caption{Selection of spectroscopic targets. The upper panel shows the selection ($g_{fiber} -
r_{fiber}$) as a function of $r_{fiber}$. The limiting $r_{fiber} = 21$ is obvious. Green dots indicate galaxies with Hectospec redshifts, magenta dots indicate cluster members, and black dots are galaxies without redshifts. The red sequence of cluster members is evident. A few galaxies fainter than the limit are from SDSS. The lower panel shows the resulting
selection in $r_{petro}$. The effective limiting apparent magnitude of the survey is  $r_{petro} = 20.5$. Red circles highlight the red sequence (the line shows the fit in equation (2)); blue crosses denote the blue cluster members. 
\label{cmr}}
\end{figure}

Figure \ref{cmr} (upper panel) shows the impact of the photometric selection on the A383 redshift survey: black dots represent all of the extended objects in the survey region, green dots are galaxies with a redshift, and magenta dots represent cluster members. There is no
$g_{fiber} -r_{fiber}$ color selection of the spectroscopic targets. The selection limit at
$r_{fiber} = 21$ is obvious.  The red sequence of cluster members is also evident. A few objects with fainter $r_{fiber}$ have redshifts from the SDSS DR10 (Ahn et al. 2013).

\begin{figure}[htb]
\centerline{\includegraphics[width=7.0in]{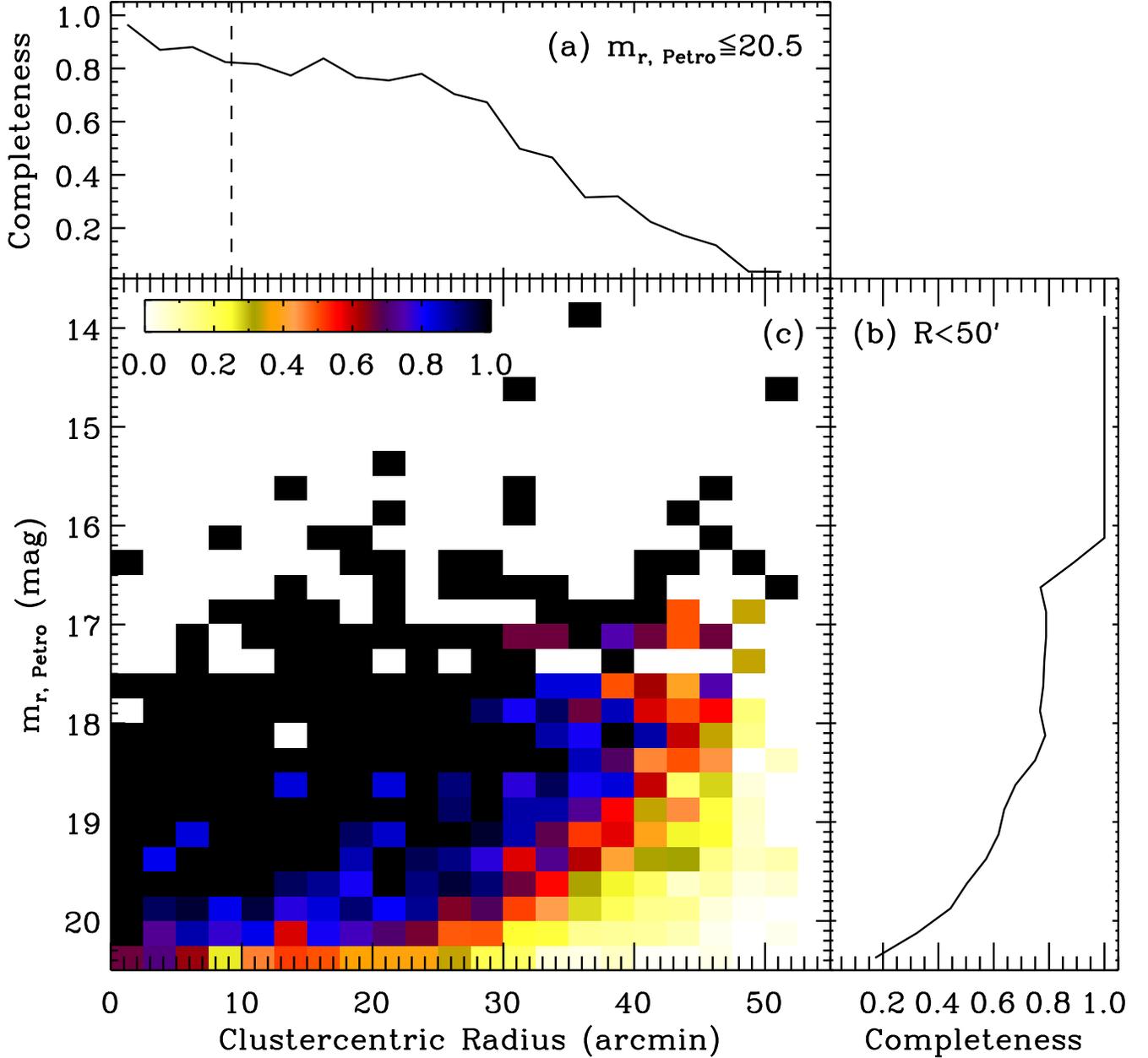}}
\vskip -2ex
\caption{Two-dimensional fractional completeness of the A383 redshift survey within 50${^\prime}$ of the BCG as a function of $r_{petro}$ and angular separation from the BCG (lower left panel). The vertical dotted line in the upper panel shows the integrated completeness as a function of radius indicates $R_{200}$. The right-hand panel show the integrated completeness as a function of $m_{r, Petro}$. 
\label{complete}}
\end{figure}

Figure \ref{cmr} (lower panel) shows the more astrophysically useful distribution of  $g_{model}-r_{model}$ as a function of $r_{petro}$. In Section \ref{members} below, we segregate the cluster red sequence (red dots) and the blue population (blue crosses). Few galaxies with $r_{petro} > 20.5$ have redshifts. Figure \ref{complete} shows the
two-dimensional distribution of the completeness of the A383 redshift survey as a function of $r_{petro}$  and angular separation from the BCG for galaxies within 55$^\prime$ of the BCG and with $r_{petro} \leq 20.5$. The survey has high, uniform completeness  within $\sim 25^{\prime}$ of the BCG.

Table 1 lists the 2360  Hectospec redshifts for galaxies in the region around A383.  The Table includes 
the SDSS ObjID (DR10), the right ascension, declination, $r_{petro}$ from the SDSS DR10,  the redshift ($z$) and its error, the redshift source, and the cluster membership (as determined from the caustic technique) flag. 

Table 1 also lists 150 redshifts from other sources. There are 137 redshifts from SDSS DR10 (Ahn et al. 2013); these redshifts are part of the BOSS project (Dawson et al. 2013) and they include only one cluster member.  Newman et al. (2013a) show a histogram based on redshifts measured with a single LRIS mask on  Keck 1 (Oke et al. 1995). Andrew Newman kindly provided this list of unpublished redshifts; only four of these are for galaxies not
in our survey (one of these four is a cluster member) and we include them here. Finally we include 7 galaxy redshifts from NED and 2 unpublished redshifts measured with the 1.5-meter telescope on Mount Hopkins. We note the sources for individual redshifts in Table 1. The total number of individual galaxy redshifts in this region is 2510.

\begin{figure}[htb]
\centerline{\includegraphics[width=7.0in]{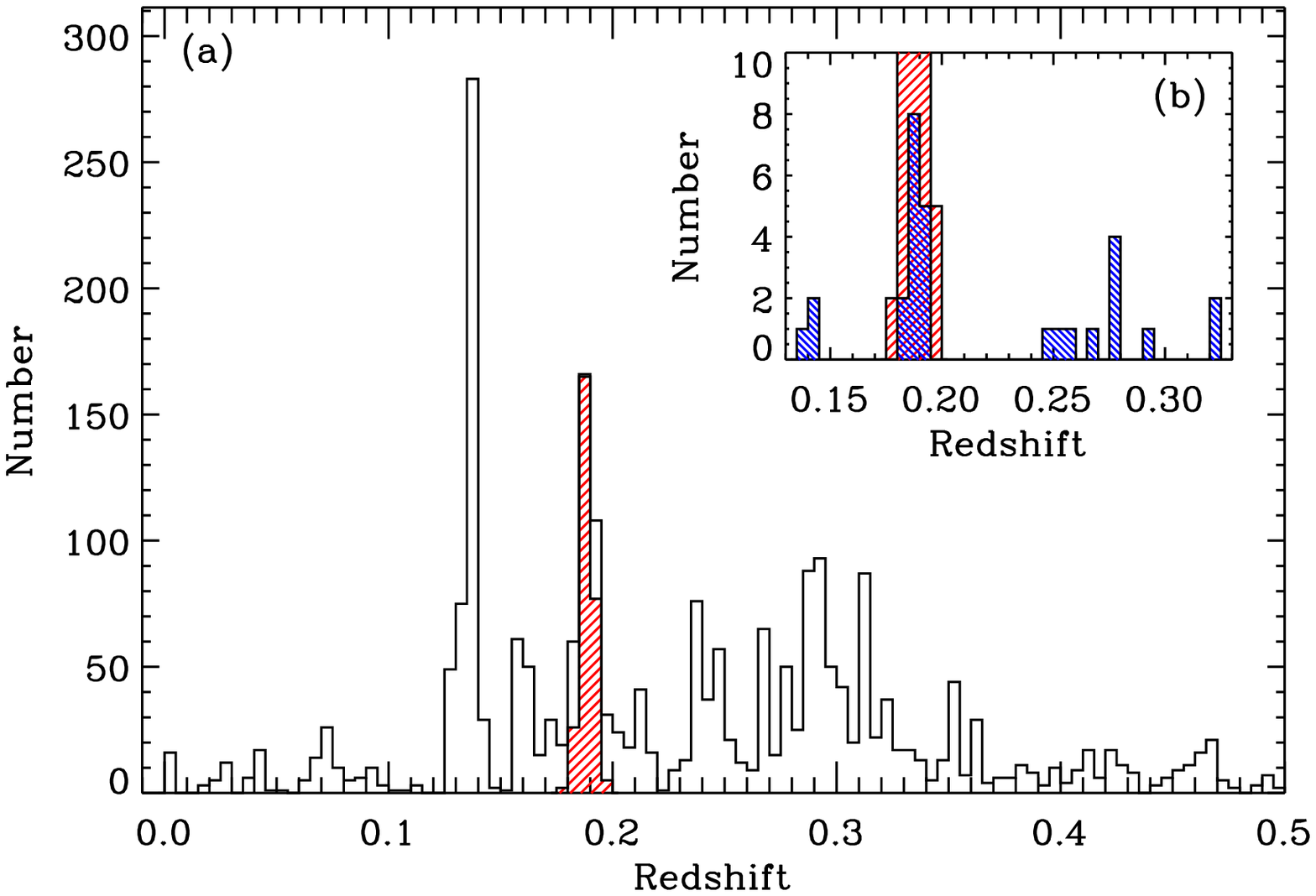}}
\vskip -2ex
\caption{Redshift distribution of galaxies in the A383 redshift survey. A383 is the peak centered at $z = 0.1887$. The hatched histogram shows the cluster members. The inset (with the same binning as the main histogram) shows the redshift distribution (blue histogram) inside the dashed circle surrounding the secondary weak lensing peak in Figure \ref
{lenover}. The red hatching in the inset indicates the redshift range for cluster members as in the full histogram.
\label{irz}}
\end{figure}

Figure \ref{irz} shows a redshift histogram for the entire survey (open histogram). The cluster A383 has a mean redshift $ z = 0.1887$ (see Section \ref{members}) and the corresponding peak in the histogram is obvious. The largest  peak is a foreground structure containing several groups of galaxies. It is interesting that a red sequence corresponding to this structure is evident in Figure \ref{cmr}; this red sequence is bluer than the red sequence of the cluster itself and overlaps it.

\begin{figure}[htb]
\centerline{\includegraphics[width=7.0in]{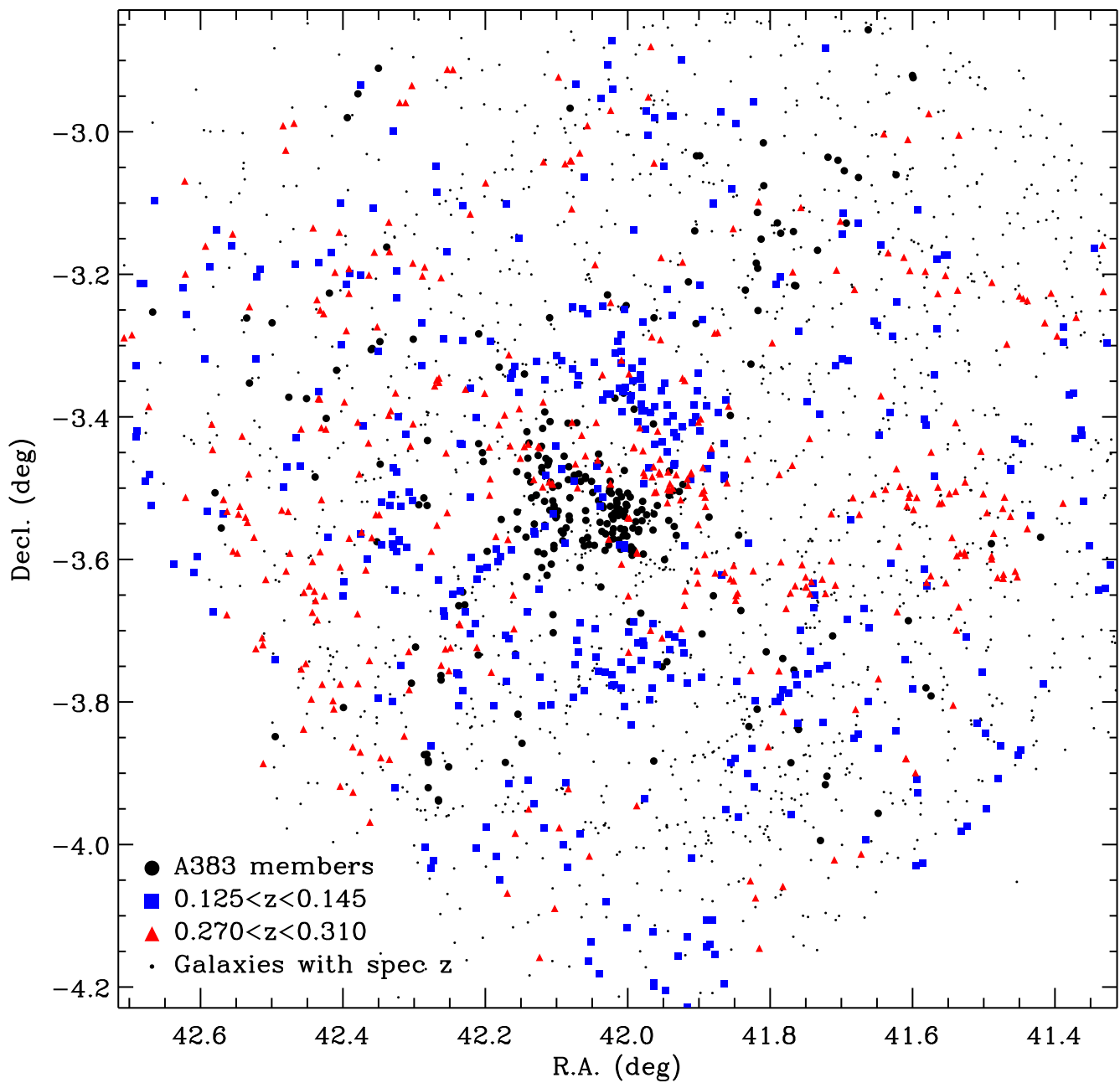}}
\vskip -2ex
\caption{ Distribution on the sky of galaxies with A383 survey redshifts. Large black dots denote cluster members; blue squares denote galaxies in the foreground peak with redshift $0.125 < z < 0.145$. Red triangles denote background galaxies with $0.270 < z < 0.310$. Small black dots mark the positions of other galaxies with spectroscopic redshifts.
\label{allz}}
\end{figure}

Figure \ref{allz} shows the distribution of galaxies with measured redshifts on the sky. The large black dots indicate the cluster members (see Section \ref{members}). The blue squares show galaxies in the foreground peak. Clumps corresponding to groups of galaxies in this redshift range are evident. Red triangles denote background objects in the redshift range $0.27 < z < 0.31$. Small black dots indicate galaxies with a redshift outside the
highlighted ranges. There are dense knots of both foreground and background objects; fortunately none of these are superimposed directly on the central region of A383. We discuss this issue further in
Section \ref{weakmap}.

\section {Cluster Membership: An Application of the Caustic Technique}
\label {members}

The caustic technique was originally derived to estimate the mass profiles of clusters of galaxies
(Diaferio \& Geller 1997 (DG97); Diaferio 1999 (D99); Serra et al. 2011). Gifford \& Miller (2013) and Gifford et al. (2013) explore variants of the technique.

The caustic technique takes advantage of the appearance of clusters in redshift space.
Clusters of galaxies appear as well-defined trumpet-shaped patterns in the phase-space defined by redshift and projected radius from the cluster center (Kaiser 1987; Regos \& Geller 1989; DG 97; D99).
DG97 first showed that the sharp boundary that delineates this pattern can be identified with the escape velocity from the cluster thus providing a route to estimation of the mass profile of the cluster  on scales up to 5-10h$^{-1}$ Mpc. 

Because the caustic technique treats each galaxy as a massless tracer of the velocity field, application of the technique is insensitive to the survey completeness. It is sensitive to the number of galaxies in the sample (Serra \& Diaferio 2013). 
The most serious limitation of the caustic technique (and of many other mass estimators) is the assumption of spherical symmetry. Serra et al (2011) show that projection effects produce a 1$\sigma$
uncertainty of 50\% in the mass profile within the radial range (0.6-4)R$_{200}$.

Serra \& Diaferio (2013) show that the original technique is  an excellent tool for separating the cluster members from the foreground and background.
They apply the technique to mock catalogs for 100  clusters extracted from N-body simulations. The cluster selection is independent of their dynamical state and morphology. This approach identifies at least 95$\pm$ 3\%  of the true cluster members within 3R$_{200}$. Only $\sim 2$\% of the galaxies within R$_{200}$ are foreground or background galaxies; the fraction of interlopers reaches $\sim 8$\% within 3R$_{200}$.

We next review this application of the caustic technique following DG97  and D99. In Section
\ref{weakmap} we display the distribution of cluster members superimposed on the weak lensing map of A383 published by Okabe et al (2010) and discuss the implications of the comparison.

The first step in applying the caustic technique is identification of the cluster center. We isolate the cluster initially by selecting all of the galaxies in our redshift survey that lie within  10$h^{-1}$ Mpc and 5000 km s$^{-1}$ of the nominal x-ray cluster center. We then construct a binary
tree based on pairwise estimated binding energies. We use the tree to identify the largest 
cluster in the field and we adaptively smooth the distribution of galaxies within this cluster to identify its center (see D99 and Serra et al. 2011) for detailed descriptions of this
process).

\begin{figure}[htb]
\centerline{\includegraphics[width=7.0in]{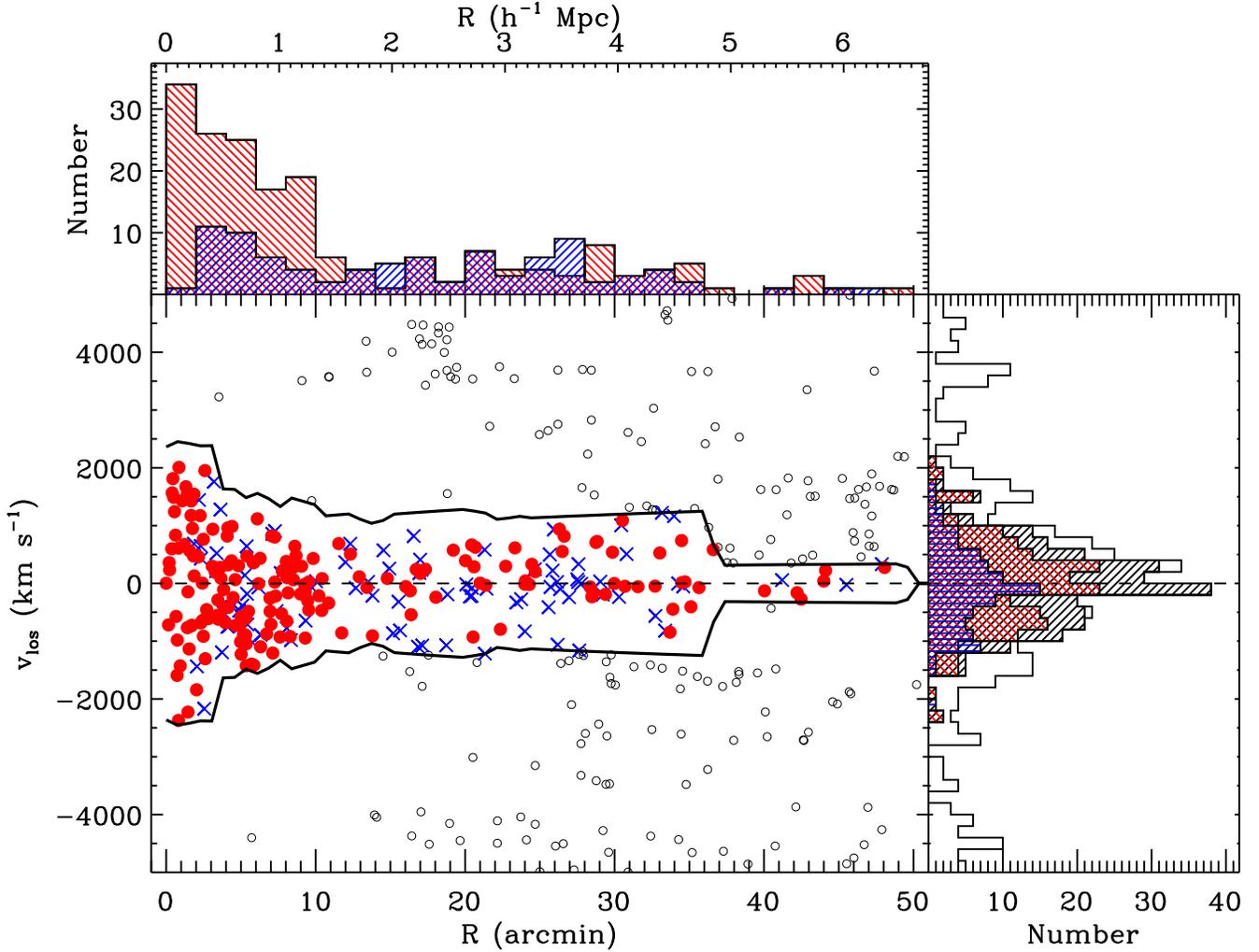}}
\vskip -2ex
\caption{Redshift diagram for A383. In redshift space the cluster is centered on the BCG (very close to the hierarchical cluster center). The vertical axis is the rest-frame velocity of the galaxy (black point) with respect to the cluster center and the horizontal axis is the projected distance from the center.
The black lines are the caustics. The median error in the amplitude of the caustics is 17 km s$^{-1}$; the error is small because the cluster is well-isolated in redshift space. We define cluster members as galaxies lying within the caustics (red dots are red sequence members; blue crosses are blue members). In the right-hand plot, the gray hashed histogram shows the
rest-frame line-of-sight velocity distribution for cluster members; the red histogram shows the red sequence cluster members and the blue histogram shows the blue members. The top red and blue histograms show the radial distribution of red and blue cluster members, repectively. 
\label{reddiag}}
\end{figure}

Once we have identified a center we can plot the distribution of galaxies in azimuthally summed phase space. This effective azimuthal averaging smooths over small-scale substructure particularly at large projected radius. Figure \ref {reddiag}  shows the distribution of galaxies in the rest frame line-of-sight velocity versus projected spatial separation plane for A383. The expected trumpet-shaped pattern centered on the mean cluster velocity is evident. 

To measure the amplitude $A(r)$ of the phase space signature of the cluster, we smooth the 
patterns  in Figure \ref{reddiag}  and identify a threshold in phase-space density as the edge of the caustic envelope. 
We define the threshold $\kappa$ by solving the equation 
\begin{equation}
|\langle{v_{esc}}^2\rangle_{\kappa,R} -\langle{v^2}\rangle_R|=0
\end{equation}
where   $v_{esc}$
is the escape velocity at radius $R$ (D99
and Serra et al. 2011). The values of the upper caustic, $A^+(r)$, and lower caustic, $A^-(r)$, 
can differ; the smaller of these two values provides our estimate of the caustic amplitude, $A(r)$.
Especially at small radii, the boundaries  of A383 (Figure \ref{reddiag}) are impressively clean and correspond well with limits one might draw by eye. 

The estimate of the error in the position of the caustics  depends
on the ratio of the number of galaxies within the caustics and the total number of galaxies at a particular radius (see Section 5.5 of Serra et al. 2011). In general, the caustics we observe in the real universe are more sharply defined than in simulations. The caustic location  depends on
the size of the galaxy sample. For samples ranging from 200-350 members (similar to A383), Serra et al. (2011) show that  the caustic amplitude is unbiased.
Because the well-sampled redshift diagram for A383
is so clean and the trumpet shape is thus so well-defined within 2$h^{-1}$ Mpc, the median error in the caustic amplitude is a negligible
17 km s$^{-1}$.  

Figure \ref{reddiag} shows the 275 member galaxies (red dots indicate red sequence members; blue crosses indicate blue members)   within the A383 caustics.   We define the red sequence
(Figure \ref{cmr}) by fitting
the member galaxies with $ 16<r_{Petro}<19.5$ and $1<(g_{model}-r_{model})<1.5$  with the linear relation 
\begin{equation}
(g_{model}-r_{model}) = (-0.027 \pm 0.011) r_{Petro} + (1.72 \pm 0.21) ~ .
\end{equation}
The rms scatter around this relation is 0.08 mag. We separate the red and blue populations with 
equation (2) moved 3$\sigma$ to the blue.  This definition is consistent with the general approach taken in other investigations of cluster galaxy populations (e.g. S{\'a}nchez-Bl{\'a}zquez et al. 2009; Rines et al. 2013). With this of the red sequence, there are 185 red members and 90 blue members.

A biased selection in galaxy color might, in principle, affect the determination of the caustic amplitude.The upper panel in Figure \ref{reddiag} shows the number of red and blue cluster members as a function of radius. It is clear from this figure that the red population dominates within within 1.5h$^{-1}$ Mpc. Thus, as Rines et al. (2013) show, the caustics are unbiased provided that the selection contains a dense enough, sufficiently broadly defined region around the red sequence.  

Galaxies with different luminosity could also populate different
regions of the velocity field. Wu et al. (2013) and Saro et al. (2013) demonstrate that with redshifts
for $\gtrsim$30 cluster members within the virial radius, the velocity dispersion is unbiased. Our survey of A383 contains more than 140 cluster members within the virial radius. Furthermore, 
Rines \& Diaferio (2006) show that the caustic mass is insensitive to galaxy luminosity provided that the redshift survey probes to a limiting absolute magnitude of at least M$_r^*$ + 1; our survey of A383 reaches to M$_r \sim$M$_r^*$ + 2. 

We conclude that systematic errors resulting from any biases in the galaxy sample in A383 are very small because we have a deep, dense, nearly complete redshift survey with no color selection.
The BCG has no effect on the position of the caustics because it sits almost exactly on the cluster mean redshift.

The hierarchical center of A383 is R.A.$_{2000}$ = 42.013789$^\circ$, Decl$_{2000}$ = -3.526626$^\circ$, $z$ = 0.188717. Based on analysis of $\sim 2700$ mock clusters, Serra et al. (2011) show that the 1$\sigma$ uncertainty in the redshift of hierarchical center is 107 km s $^{-1}$ in the rest-frame of the cluster and the positional uncertainty is $\sim 56 h^{-1}$ kpc. These uncertainties dominate the uncertainty in the difference between the position of the BCG and the cluster center.    

The  projected separation between our hierarchical center and the BCG is 21$\pm 56 h^{-1}$ kpc. The BCG offsets from the x-ray center and the lensing center of the cluster are $\lesssim 3 h^{-1}$ kpc. 

The Hectospec redshift of the BCG is $ z = 0.188713$.
In the rest frame of the cluster the
line-of-sight velocity difference between the BCG and the hierarchical center is
$-11 \pm 110$ km s$^{-1}$. Newman et al. (2013a) report a consistent but less well-constrained BCG peculiar velocity of $-261 \pm 187$ km s$^{-1}$ based on 26 unpublished redshifts in the cluster.  
Thus both the BCG redshift and position are consistent with the center of the dark matter halo.

\section {The Velocity Dispersion Profile}
\label{vprofile}

In a regular, relaxed cluster where the BCG coincides with the center of the dark matter halo, the stellar velocity dispersion of the central cD galaxy may provide a measurement of the total gravitational potential (Miralda-Escude 1995; Natarajan and Kneib 1996; Newman et al 2013a).
At radii $\gtrsim$ 10 kpc the stellar velocity dispersion profile extends into the dark matter dominated region of the cluster and overlaps the velocity dispersion profile determined from a dense redshift survey of cluster member galaxies (see e.g. Kelson et al. 2002: Newman et al. 2013b (Figure 3)). Modulo the possibly differing velocity anisotropies of the stars in the cD and the galaxies in the cluster, the two tracers potentially provide complementary measures of the cluster potential.

\begin{figure}[htb]
\centerline{\includegraphics[width=7.0in]{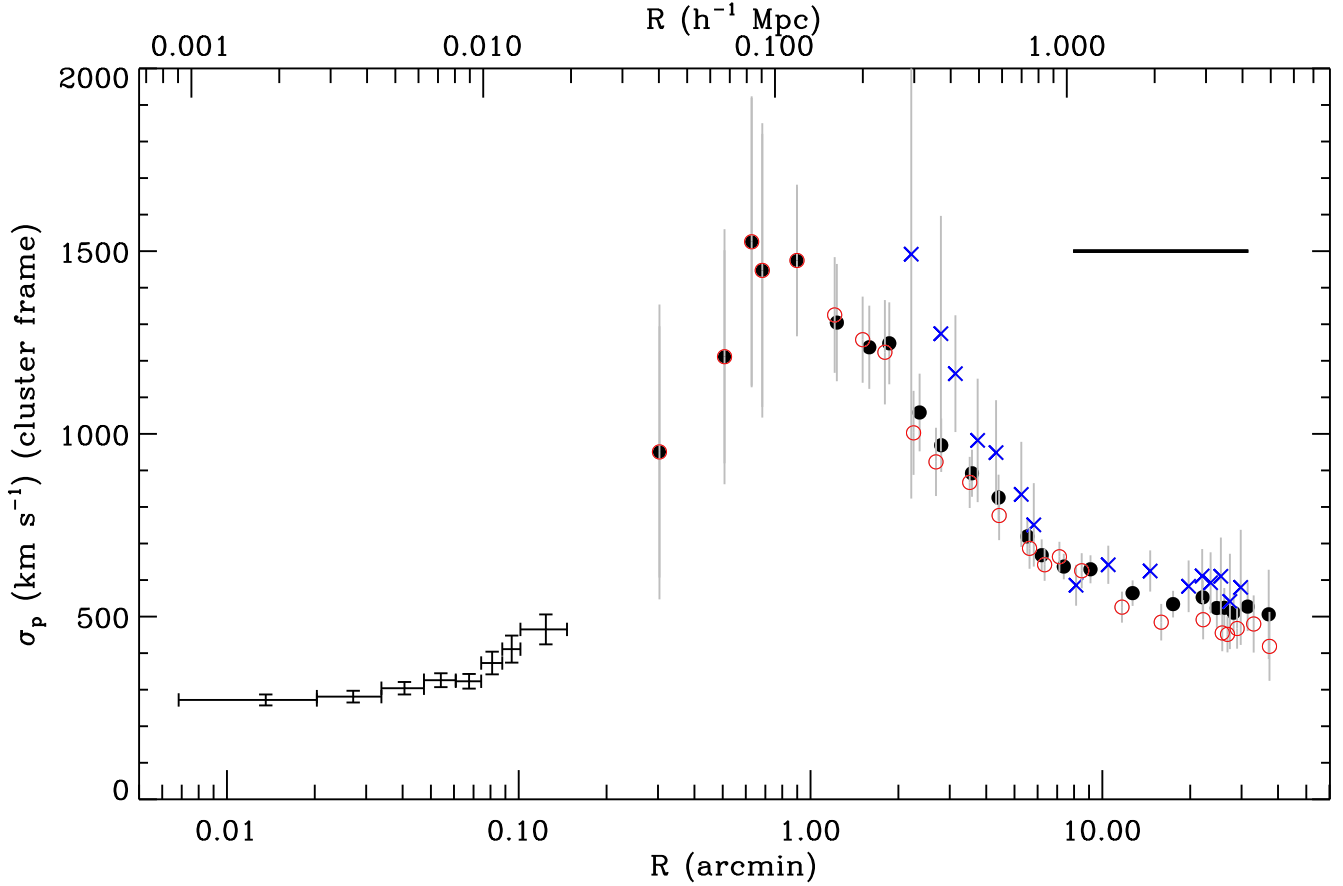}}
\vskip -2ex
\caption{Velocity dispersion profile for A383. At radii $\lesssim 0.1$ arcmin, points with 1$\sigma$ error bars show the velocity dispersion profile of the BCG from Newman et al. (2013a). At large radii we derive the velocity dispersion profile for the cluster members selected according to the caustic technique. The horizontal bar shows the extent of the overlapping logarithmic bins (0.6 dex throughout). Black points refer to the entire sample, red open circles refer to the red sequence members; blue crosses refer to
blue members. See Fig. \ref{reddiag} for the number of red/blue members as a function of radius. At small radii the number of blue members is inadequate to determine a dispersion. 
\label{vdisp}}
\end{figure}

Newman et al. (2011; 2013a) measured a velocity dispersion profile for the cD in A383 that extends to a radius of $\sim 27 h^{-1}$ kpc. Figure \ref{vdisp} shows this profile from Table 6 of Newman et al. (2013a).
Figure \ref{vdisp}  also shows the rest frame  line-of-sight velocity dispersion for all galaxies (solid black points) in the cluster. The horizontal bar indicates the extent of the overlapping logarithmic bins. The turnover of the galaxy velocity dispersion profile at 
radii $\lesssim 100 h^{-1}$ kpc suggest a consistency between the  stellar velocity dispersion and
the cluster member velocity dispersion as tracers of the potential. 

In a  pilot study of the nearby rich cluster A2199, Kelson et al. (2002) observe similar behavior.
The peak in the aggregate velocity dispersion profile is a direct consequence of density profiles of the generalized NFW (Navarro et al. 1997) form

\begin{equation}
\rho \propto {1 \over x^\alpha(x+1)^{3-\alpha}}
\end{equation}

\noindent with reasonable velocity anisotropy (see Navarro et al. 2010). Kelson et al. (2002) fit spherically symmetric models to their data. They find that no single component model with reasonable velocity anisotropy can fit the data. A two-component model that treats the stellar and dark matter components separately provides an acceptable fit. Their fit implies a soft core with $\alpha < 1$ rather than the expected cusp.
The challenges of model fits to the combined BCG and galaxy velocity dispersion profile are beyond the scope of this paper; we plan to consider this issue later in combination with other systems. 

Here we display the velocity dispersion profile and examine the impact of the blue cluster population.  Figure \ref{vdisp} shows the velocity dispersion profiles for the red sequence only (open red circles) and for the blue population (blue crosses). Within the virial radius, the blue population
has little impact on the velocity dispersion profile. At larger radii the velocity dispersion of the blue population systematically exceeds the velocity dispersion for the red population although the difference only occasionally exceeds the 1 $\sigma$ error only at radii $\gtrsim 1h^{-1}$ Mpc. 

The small impact of blue cluster members on the line-of-sight cluster velocity dispersion profile is consistent with the analysis by Rines et al. (2013) of the clusters A267, A2261, and RXJ2129. Mahajan et al. (2011) also demonstrate that the cluster velocity distribution is insensitive to color.

\section {The Cluster Member Distribution and Weak Lensing}
\label{weakmap}

\begin{figure}[htb]
\centerline{\includegraphics[width=7.0in]{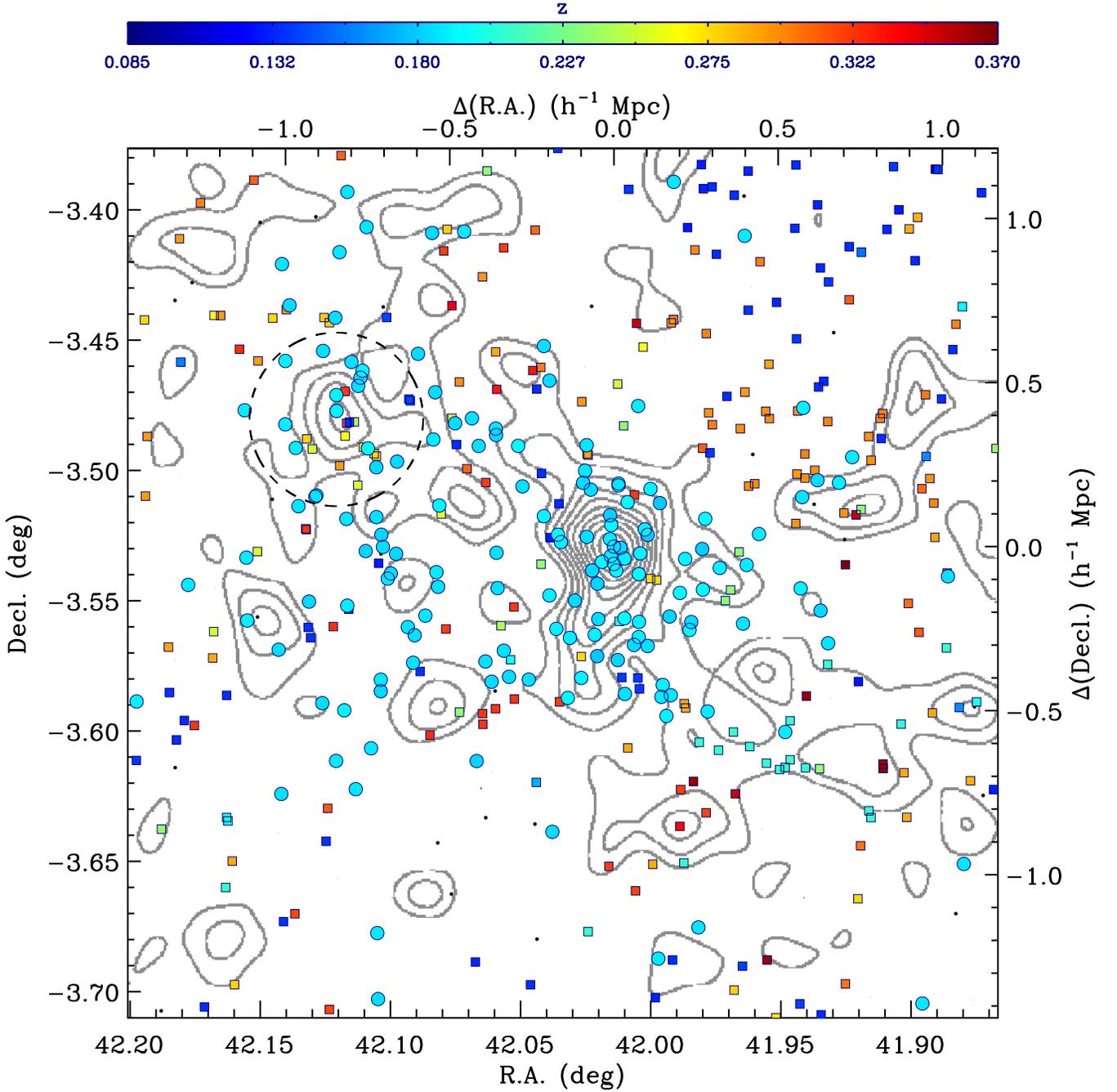}}
\vskip -2ex
\caption{Cluster members (cyan colored circles) superimposed on the weak lensing map by Okabe et al. (2010). In this region the redshift survey is $\sim 80$\% complete to $r_{petro} = 20.5$. The weak lensing contours ${\kappa}{(\Theta)}$ are spaced in $\Delta\kappa = 1 \sigma_{\kappa}$; the lowest contour is $1\sigma_{\kappa}$. See Okabe et al. (2010) for a complete
description of the derivation of the contours. Non-members (squares) are also superimposed and color-coded by redshift according to the color bar. Black dots indicate galaxies with a redshift outside the color bar redshift window. The dashed circle centered on the secondary weak lensing peak
has a 2$^\prime$ radius (264 $h^{-1}$ kpc at the mean cluster $z = 0.1887$). The inset in Figure \ref{irz} shows the redshift distribution in this region.
\label{lenover}}
\end{figure}

Okabe et al. (2010) published a weak lensing map for a 20$\times$ 20 arcminute field centered on 
the BCG of A383. In addition to A383 itself (R.A.$_{2000}$ = 42.01567$^\circ$, Decl$_{2000}$ = -3.53215$^\circ$)   (an 11.1 $\sigma$ peak), the map shows a 4.2$\sigma$ peak at R.A.$_{2000}$ = 42.12078$^\circ$, Decl$_{2000}$ =     -3.48041$^\circ$ (Smith 2012, private communication). Figure \ref{lenover} shows the contours from the Okabe et al. (2010) weak lensing map. 

Figure \ref{lenover} also shows the distribution of cluster members with spectroscopic redshifts. The distribution of cluster members is asymmetric relative to the cluster core and there  are several interesting aspects of the correspondence between the distribution of cluster members and the weak lensing map. First, the extension of the weak lensing map to the south corresponds to an extension of the distribution of cluster members; there are significantly fewer members to the north of the main concentration.  Second, cluster members extend to the east (and not to the west) of the main concentration. 

\begin{figure}[htb]
\centerline{\includegraphics[width=7.0in]{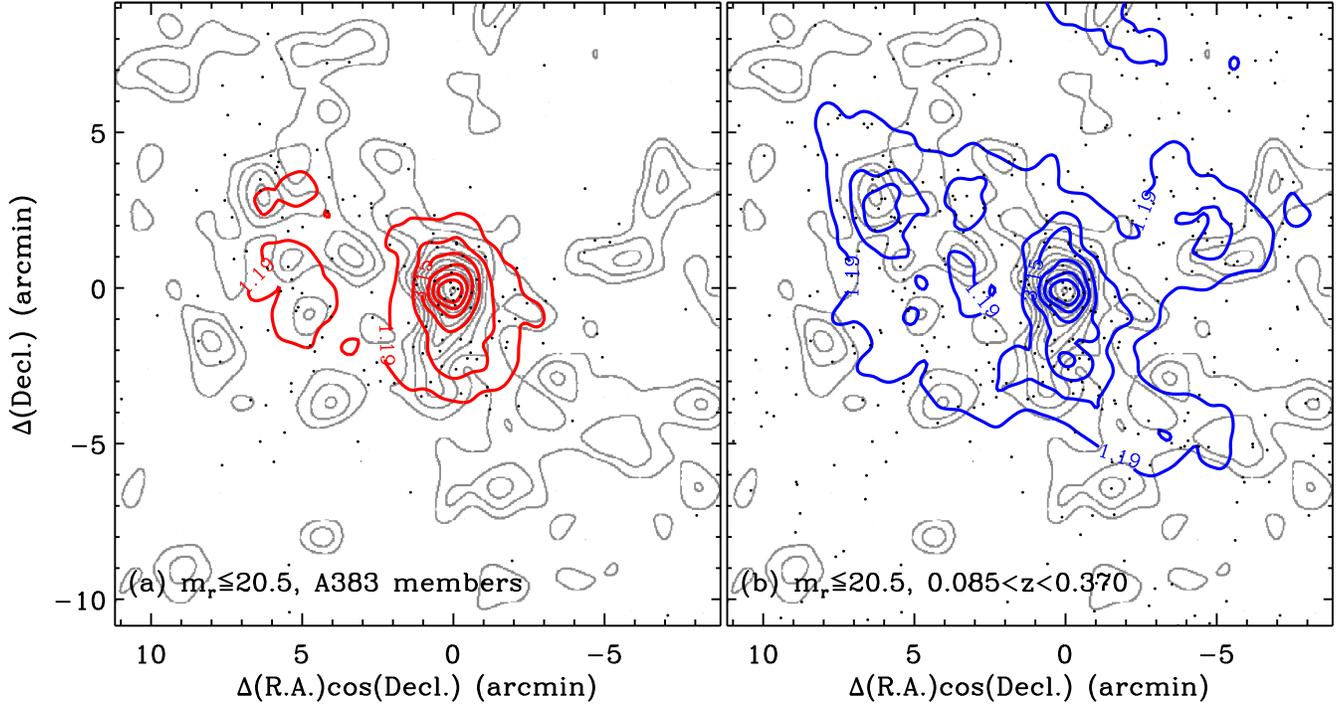}}
\vskip -2ex
\caption{Cluster member number density contours (red; left) superimposed on the weak lensing map (gray).
Black dots denote galaxies used to construct the number density contours.
The lowest  cluster surface number density contours is 1.19 galaxies arcmin$^{-2}$ and the
contours increase in steps of 
1 galaxy arcmin$^{-2}$. The secondary peak in the number density distribution that coincides with the secondary weak lensing peak is insignificant. The right panel shows similar number density contours (blue) for all galaxies in the redshift survey superimposed on the weak lensing map. The secondary peak in the number density distribution that coincides with the secondary weak lensing peak is significant at the $\sim 4.9 \sigma$ level. The plot suggests that background structure significantly enhances the significance of the secondary peak. 
\label{lensdensity}}
\end{figure}

Figure \ref{lensdensity}  (left panel) shows the smoothed surface number density distribution of cluster members (red contours). The median density of cluster members in this region is 0.15 galaxies arcmin$^{-2}$ and the 1$\sigma$ fluctuation in the number is 0.61 galaxies arcmin$^{-2}$.

The distribution of members in Figure \ref{lensdensity} differs significantly from similar plots of the possible membership distribution based on photometric redshifts published by Okabe et al. (2010) and by Newman et al. (2013a). The differences appear to result primarily from inclusion of foreground structures within the
photometric redshift membership window. 

Figure \ref{lensdensity} shows that the number density of cluster members 
concentrates on the weak lensing peak with a corresponding north-south elongation. The lowest level contour corresponds to 1.7 $\sigma$ (1.19 galaxies arcmin$^{-2}$);  thus the outlying structure to the east overlapping the secondary weak lensing peak is of very low significance. The qualitative appearance of the member contours is similar if they are luminosity weighted.

Mass condensations over a broad redshift range may contaminate the weak lensing signal (see Hoekstra et al. 2013 and references therein). At the relatively low redshift of A383, this cosmic noise is an important limitation of the determination of weak lensing mass profiles particularly at large radii
(e.g. Hoekstra 2001; Hoekstra 2003; Hoekstra et al. 2013). 
Structures with  angular size distances ranging from roughly half to twice the angular size distance to A383 have the greatest potential for contaminating the weak lensing signal. This range of angular size distances corresponds to a redshift range $0.085< z < 0.52$. Our redshift survey is very sparse for redshifts $\gtrsim 0.37$. We thus explore the potential impact of the structure we observe in the redshift range $0.085 < z < 0.37$ on the weak lensing map. We compare the smoothed galaxy number density with the lensing map keeping in mind that there may be structures in the range $0.37 < z < 0.52$ that also affect the interpretation of the map.

The right panel of Figure \ref{lensdensity} shows surface number density contours for galaxies with measured redshifts in the range $0.085 < z < 0.37$ (the contours are insensitive to inclusion of the few galaxies outside this range). The qualitative appearance of these contours is similar
if they are weighted by luminosity. The median number density of galaxies in this range and region 
is  0.71 galaxies arcmin$^{-2}$; the 1$\sigma$ variation in the number density is also 0.71 galaxies
arcmin$^{-2}$. 

There is a remarkable coincidence of a secondary peak in the galaxy number density (the significance is $\sim 4.9 \sigma$) with the secondary weak lensing peak at R.A.$_{2000}$ = 42.12078$^\circ$, 
Decl$_{2000}$= -3.48041$^\circ$. This coincidence suggests that the secondary lensing peak results from a superposition of the outskirts of A383 with foreground/background structures at a range of redshifts. 

This analysis suggests that weak lensing mass profiles
could be improved (even near the virial radius)  by using a redshift survey to identify
structures superposed along the line of sight and within the
lensing kernel (see, for example, Coe et al. 2012). Here we estimate that the effective mass projected on the secondary peak is $\sim 10^{13}h^{-1}$ M$_\odot$, insufficient to affect either the 
cumulative caustic mass (see Section \ref {massprof}) or cumulative weak lensing mass within R$_{200}$ significantly. Geller et al. (2013) highlight the detectable impact of two clusters, Abell 750 and MS0906+11, superimposed  along the line-of-sight on the weak lensing mass. More extensive comparisons of the relevant galaxy (light) distributions determined from a dense redshift survey with weak lensing maps would provide important limits on the direct detectability of foreground/background structures impacting mass estimates for individual systems even within R$_{200}$.

\section {Caustic and Weak Lensing Mass Profiles}
\label{massprof}

Geller at al. (2013) show that near the virial radius the lensing and caustic method agree remarkably well for a sample of 19 clusters. As a result of increasingly  large, well-defined redshift surveys of individual clusters, the caustic technique has  come into wide use (e.g. Biviano et al. 2013; Lemze et al. 2013; Munari et al. 2013). 

The weak lensing profile for A383 extends to $\sim 1.2$R$_{200}$; the extent is limited by the size of the image.   
Here we compare the caustic and weak lensing profiles for A383 focusing on the region around R$_{200}$. We restrict comparison of mass estimates to these two techniques because neither depends on an equilibrium assumption and both yield
a reasonably robust, unbiased measure of mass profile at large radius (for a recent discussion of weak lensing see Hoekstra et al. 2013).  

We apply the caustic technique as originally developed by DG97 and D99  to the redshift survey data. 
DG97 show that the mass of a spherical shell within the infall region is the integral of the square of the caustic amplitude $A(r)$:
\begin{equation}
GM(<R) - GM(<R_0) = F_\beta \int_{R_0}^{R} A^2(x)dx
\end{equation}

\noindent where $F_\beta \simeq 0.5$ is a filling factor estimated from numerical simulations (D99). We approximate $F_\beta$ as a constant. Serra et al. (2011) take $F_\beta$ = 0.7; this value compensates for  the difference between D99 and Serra et al. (2011) in the identification of the ``sigma plateau'' for cutting the
binary tree. The D99 and Serra et al. (2011) approaches yield identical mass profiles to within the errors. Variations in $F_\beta$ with radius lead to some systematic uncertainty in the mass profile we derive from the caustic technique.
Serra et al. 2011 show that, within 0.6R$_{200}$, the caustic method overestimates the cluster mass by $\sim70$\% on average. This overestimate is a consequence of the radial dependence of the filling factor F$_\beta$. For an individual system, the overestimation of the central mass propagates into an overestimate of the concentration.
At $R_{200}$ the bias becomes negligible.

The caustic technique yields a direct estimate of the characteristic radius $R_{200}$ (Table \ref{tbl:props}). Within this radius there are N$_{200,obs}$ = 140 member galaxies with measured redshifts brighter than $r_{petro} = 20.5$. We correct this  
estimate for incompleteness according to the two-dimensional  map in Figure \ref {complete}.
We assume that in every pixel the membership fraction among the unobserved galaxies is the same as for those observed; the estimated membership within $R_{200}$ is then N$_{200, corr}$ = 157$\pm$4. 

Table \ref {tbl:props} also lists the parameters of the NFW fit to the caustic mass profile within 
1 h$^{-1}$ Mpc. 
We derive a concentration c$_{200} = 14.1 \pm 0.4$ and Newman et al. (2013a), for example derive c$_{200} = 6.51^{+0.92}_{-0.81}$. This difference once again highlights the known systematics in the caustic mass relative to the lensing mass on small scales (see Geller et al. 2013). 

\begin{figure}[htb]
\centerline{\includegraphics[width=7.0in]{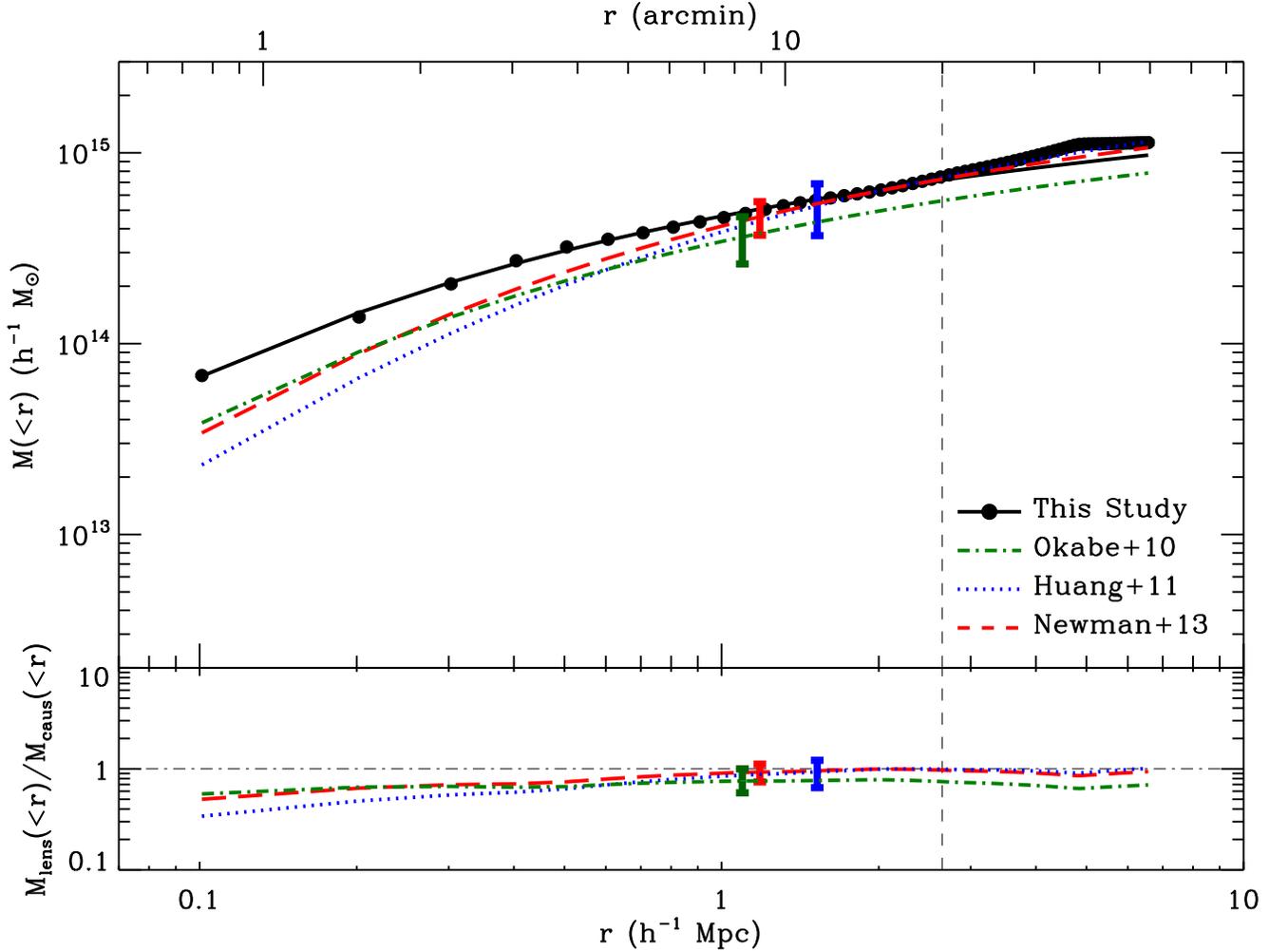}}
\vskip -2ex
\caption{Comparison of cumulative mass profiles derived from weak lensing (dotted lines) and the caustic technique
(points; solid line is the NFW fit). The weak lensing profiles are from Okabe et al (2010; Table 6), Huang et al. (2011; Table 4, Model d),  and Newman et al. (2013a; Table 8, line 3). The lower panel shows the ratio between the weak lensing and caustic profiles. Error bars for the lensing profiles indicate the error at the virial radius (Okabe et al 2010; Huang et al. 2011) or at $R_{200}$ (Newman et al. 2013). The point size for the caustic mass indicates the error at $R_{200}$. These 1$\sigma$ errors are
are reasonably representative of the errors throughout the radial range. The agreement among the profiles is excellent for R $\gtrsim 0.8 R_{200}$; at smaller radii the caustic mass profile exceeds the weak lensing profile as expected (see Geller et al. 2013). The vertical dotted line indicates the 
radial limit of the weak lensing data; only the dynamical data provide a direct measure of the mass profile at larger radii.
\label{fig:A383mass}}
\end{figure}

Figure \ref{fig:A383mass} shows the caustic mass profile along with three recent lensing mass profiles. Although there is some overlap in the data
underlying the lensing analyses, the approaches differ. Okabe et al. (2010) analyze Subaru i$^{\prime}$ and V-band images. They identify
background galaxies by constructing a sample of galaxies both bluer and redder than the cluster red sequence. They choose offsets from the red sequence for these source galaxies by minimizing the apparent dilution of the lensing signal by possible cluster members. Although their NFW (Navarro, Frenk \& White 1997) model does not fit the weak lensing data well, Okabe et al (2010) do list representative parameters in their  Table 6. We show the profile as a green dot-dashed line in Figure \ref{fig:A383mass}.  Huang et al. (2011) base their mass profile (Table 4, Model d) on a more extensive set of imaging data. They include $BVRIz$ images from Subaru along with $u$ imaging from CFHT MEGACAM.  From this wealth of imaging data, Huang et al (2011) select sources on the basis of photometric redshifts for their favored weak lensing mass estimate (Method d in their Table 4; blue dotted line in Figure \ref{fig:A383mass}).  Newman et al. (2013a) base their profile on a combination of strong lensing, weak lensing data and observation of the velocity dispersion profile of the BCG. In this way they constrain the total density profile for the clusters in their sample over three decades in radius. We show one of their fits
(Table 8, line 3) to the A383 data as a red dashed line in Figure \ref{fig:A383mass}.

The lower panel of Figure \ref{fig:A383mass} shows the ratio between the caustic lensing mass profiles.
The comparison highlights discrepancies at small radii and impressive agreement for $R \gtrsim 0.8R_{200}$ for the Huang et al. (2011) and Newman et al. (2013a) profiles. The Okabe et al. (2010) NFW fit may agree less well as a result of more limited photometric data. The caustic mass profile is completely independent of the sample of lensing profiles; thus the agreement with the most recent profiles at large radii is impressive. The error bars in Figure \ref{fig:A383mass} show the quoted 1$\sigma$ error at $R_{200}$ (Newman et al. 2013a) or $R_{vir}$ (Okabe et al. 2010: Huang et al. 2011). For the caustic mass profile, the error near the virial radius is comparable with the size of the dots. The caustic mass  and the lensing masses are clearly consistent with one another at these radii. At larger radius the extrapolations of the NFW fits to the Newman et al. (2013a) and Huang et al. (2012) lensing profiles also agree well with the independent, direct caustic estimate.  

The mass profiles measured by weak lensing and by a combination of strong and weak lensing have the highest confidence at small radius $\lesssim R_{200}$. It is thus reasonable to regard the lensing profiles in this regime as the true profile. Then, as emphasized by Geller et al (2013), the comparison provides a measure of the systematic introduced by the assumption of a constant
filling factor, F($\beta$). 

Table \ref{tbl:props} lists the dynamical parameters of A383 derived from the redshift survey.
The value we derive for $R_{200}$  agrees very well with the lensing value, $R_{200}$ = 1.183$^{+0.089}_{-0.071}$ $h^{-1}$ Mpc (Newman et al. 2013a). The rest frame line-of-sight velocity dispersion within R$_{200}$ is $931 \pm 59$ km s$^{-1}$, consistent with the value Okabe et al (2010) obtained for their SIS (singular isothermal sphere) fit to the weak lensing data (875$^{+34}_{-31}$ km s$^{-1}$; their Table 4). The caustic mass we obtain for M$_{200}$ also agrees with the
Newman et al. (2013a) value,  M$_{200} = 4.6 ^{+1.1}_{-0.8} \times 10^{14} h^{-1}$ M$_\odot$.

\section {Conclusion}

Robust mass estimates for rich clusters of galaxies are part of the foundation for testing  models of structure formation in the universe. Weak lensing and the caustic technique based on large redshift surveys of individual clusters both reach to the virial radius. Both techniques are independent of equilibrium assumptions. These mass estimation techniques provide completely independent
measures of the mass within the fiducial radius, R$_{200}$. Based on a large redshift survey (2510 galaxies) of  the region within 50$^\prime$ of the center of the strong-lensing cluster Abell 383, we derive M$_{200}$, R$_{200}$,
and the behavior of the mass profile in the infall region. All of these measures are in excellent agreement with recent independent lensing results (Okabe et al. 2010; Huang et al. 2012; Newman et al. 2013a).

Based on the caustic technique (Serra \& Diaferio 2013), we identify 275 cluster members within 7$h^{-1}$ Mpc of the hierarchical cluster center. Combining these  with the velocity dispersion profile for the BCG (Newman et al. 2011; 2013a) we demonstrate that the velocity dispersion profile for cluster galaxies appears to be a natural extension of the BCG profile near the core. The velocity dispersion of cluster members is insensitive to color for radii $\lesssim  1h^{-1}$ Mpc.

Several groups (e.g. Okabe et al. 2010; Newman et al. 2013a) have compared the projected mass distribution for A383 as revealed by weak lensing with the estimated light distribution for cluster members identified with photometric redshifts. We compare the distribution of cluster members with the
weak lensing map and demonstrate the similarity in morphology. We also show that superposed 
foreground and background structure  contributes to a secondary weak lensing peak.
In principle, a dense redshift survey can be a tool for removing contamination from a lensing map. 
If systematic bias introduced by contaminating superposed structures can be reduced,  
the comparison of kinematic and lensing mass profiles at large radius could be a route to testing alternative
theories of gravity (Lam et al. 2012; Lam et al. 2013; Zu et al. 2013).

Our survey of Abell 383 adds to the growing list of massive systems where comparison of dynamical and lensing mass estimates is possible. Similar comparisons for less massive systems and for systems covering a broader range of redshifts are important for developing methods to control systematic issues in these fundamental measures.

\begin{acknowledgments}
We thank Andrew Newman for providing his unpublished redshifts and we thank Graham Smith for providing a coordinate for the position of the secondary peak in the Subaru weak lensing map. Discussions with Ian Dell'Antonio clarified many issues in the application of weak gravitational lensing. We also profited from valuable comments by Adi Zitrin and Scott Kenyon.  Perry Berlind and Michael Calkins masterfully operated the Hectospec on the MMT and Susan Tokarz efficiently reduced the data. The Smithsonian Institution supports the research of MJG, HSH, and MJK. AD acknowledges partial support from the INFN grant Indark and from the grand Progetti di Ateneo TO\_Call\_2012\_0011
`Marco Polo' of the University of Torino. 
KR is partially supported by a Cottrell College Science Award from the Research Corporation.
\end{acknowledgments}

{\it Facilities:}\facility {MMT(Hectospec)}

\clearpage
\begin{deluxetable}{rcccccccc}
\tabletypesize{\footnotesize}
\tablewidth{0pc} 
\tablecaption{Redshifts in the field of A383
\label{tab-samp}}
\tablehead{
ID & SDSS ObjID (DR10) & R.A.$_{2000}$ & Decl.$_{2000}$ & $r_{\rm Petro}$ (mag)     & $z$ & $z$ Source\tablenotemark{b} 
& Member\tablenotemark{c}
}
\startdata
   1 &    1237679323937833129 &  2:44:48.44 &$-3$:33:14.14 &  18.909 & $0.30322\pm0.00007$ &  1 &  0 \\
   2 &    1237679323937833329 &  2:44:51.95 &$-3$:35:01.97 &  19.670 & $0.17180\pm0.00010$ &  1 &  0 \\
   3 &    1237679255211344627 &  2:44:52.48 &$-3$:14:34.92 &  21.209 & $0.66462\pm0.00016$ &  2 &  0 \\
   4 &    1237679255211344283 &  2:44:53.80 &$-3$:24:35.42 &  19.254 & $0.31100\pm0.00013$ &  1 &  0 \\
   5 &    1237679323937833356 &  2:44:54.34 &$-3$:31:25.68 &  19.392 & $0.14067\pm0.00009$ &  1 &  0 \\
   6 &    1237679323937833131 &  2:44:56.32 &$-3$:30:27.56 &  18.366 & $0.17077\pm0.00012$ &  1 &  0 \\
   7 &    1237679254674473240 &  2:44:56.46 &$-3$:40:23.43 &  19.636 & $0.16156\pm0.00007$ &  1 &  0 \\
   8 &    1237673701818630486 &  2:44:57.76 &$-3$:36:43.42 &  19.668 & $0.38204\pm0.00012$ &  1 &  0 \\
   9 &    1237679323937833141 &  2:44:58.58 &$-3$:27:00.66 &  20.820 & $0.57022\pm0.00011$ &  2 &  0 \\
  10 &    1237679254674473193 &  2:44:58.71 &$-3$:39:42.23 &  18.872 & $0.32922\pm0.00009$ &  1 &  0 \\
\enddata
\tablenotetext{a}{This table is available in its entirety in a machine-readable form in the online journal. A portion is
 shown here for guidance regarding its form and content.}
\tablenotetext{b}{(1) This study; (2) SDSS DR10; (3) Newman et al. (2013a); (4) NED; (5) FLWO 1.5-meter.}
\tablenotetext{c}{(0) A383 non-members; (1) A383 members.}
\end{deluxetable}

\clearpage
\begin{deluxetable}{ccc}
\tabletypesize{\footnotesize}
\tablewidth{0pc} 
\tablecaption{Derived Properties of A383
\label{tbl:props}}
\tablehead{
Parameter & Caustics & NFW Fit
}
\startdata
   R.A.$_{2000}$ & 42.013789$^\circ$ & \nodata\\
   Decl.$_{2000}$ & $-3$.526626$^\circ$& \nodata\\
   z & 0.188717  & \nodata\\
   $\Delta$R$_{BCG}$ ($h^{-1}$ Mpc) & 21$\pm$56 & \nodata \\
   $\Delta${z}$_{BCG}$ (km/s)& $-11 \pm 110$  & \nodata\\
   
   N$_{mem}$ &  275\tablenotemark{a}    & \nodata\\
   
   R$_{200}$ ($h^{-1}$ Mpc) &  $1.220\pm 0.010$ &$1.227\pm 0.012$\\
   $\sigma_{los}$ (km s$^{-1}$) &$931 \pm 59$& \nodata \\ 
   N$_{200,obs}$ &  140\tablenotemark{b} & \nodata\\
   N$_{200 ,corr}$& $157\pm 4$\tablenotemark{c} &\nodata\\
   M$_{200}$ ($10^{14}$ $h^{-1}$ M$_\odot$) &  $5.07 \pm 0.09$&$4.29 \pm 0.14$ \\
   c$_{200}$ = $R_{200}/R_s$ & \nodata &$14.1\pm0.4$\tablenotemark{d}\\
   $\chi^2$/dof&\nodata & 1.53\\
\enddata
\tablenotetext{a}{Count refers to all galaxies within the caustics regardless of apparent magnitude.}
\tablenotetext{b}{Count of members with measured redshifts, brighter than $r = 20.5$ inside R$_{200}$.}
\tablenotetext{c}{Corrected count assuming that the membership fraction in every pixel of the completeness map (Figure \ref{complete}) is the same for observed and unobserved galaxies.}
\tablenotetext{d}{Caustic estimate biased high as discussed in Section \ref{massprof}.}
\end{deluxetable}
\clearpage

\end{document}